*Thermotropic reentrant isotropy and antiferroelectricity in the ferroelectric nematic material RM734*

Xi Chen[1], Min Shuai[1], Bingchen Zhong[1], Vikina Martinez[1], Eva Korblova[2], Matthew A. Glaser[1], Joseph E. Maclennan[1], David M. Walba[2], Noel A. Clark[1]*

[1]*Department of Physics and Soft Materials Research Center, University of Colorado, Boulder, CO 80309, USA*

[2]*Department of Chemistry and Soft Materials Research Center, University of Colorado, Boulder, CO 80309, USA*

*Abstract*

We report a transition from the ferroelectric nematic liquid crystal ($N_F$) phase to a lower-temperature, antiferroelectric fluid phase having reentrant isotropic symmetry ($I_A$), in the liquid crystal compound RM734 doped with small concentrations of the ionic liquids BMIM or EMIM. Even a trace amount of ionic liquid dopant facilitates the kinetic pathway for the transition from the $N_F$ to the $I_A$, enabling simple cooling to produce this isotropic fluid phase rather than resulting in crystallization. The $I_A$ was also obtained in the absence of specific ionic liquid doping by appropriate temperature cycling in three distinct, as-synthesized-and-purified batches of RM734, two commercial and one from our laboratory. An additional birefringent, lamellar-modulated, antiferroelectric phase with the director parallel to the layers, resembling the smectic $Z_A$, is found between the paraelectric and ferroelectric nematic phases in RM734/BMIM mixtures.

*INTRODUCTION*

Proper ferroelectricity in liquids was predicted in the 1910's by P. Debye [1] and M. Born [2], who applied the Langevin-Weiss model of ferromagnetism to propose a liquid-state phase change in which the ordering transition is a spontaneous polar orientation of molecular electric dipoles. A century later, in 2017, two groups independently reported, in addition to the typical nematic (N) phase, novel nematic phases in strongly dipolar molecules, the "splay nematic" in the molecule RM734 [3,4,5] and a "ferroelectric-like nematic" phase in the molecule DIO [6]. These nematic phases were subsequently demonstrated to be ferroelectric in RM734 [7] and in DIO [8,9], and to be the same phase in these two materials [9]. This phase, the ferroelectric nematic ($N_F$), is a uniaxially symmetric, spatially homogeneous liquid having nearly saturated polar ordering of its longitudinal molecular dipoles (polar order parameter > 0.9) [7,9]. Chiral doping gives director twist and the second new phase of the ferroelectric nematic realm, the twisted nematic helielectric polarization state ($N_F^*$) [10,11,12,13,14]. The third new phase, the antiferroelectric $SmZ_A$, exhibiting smectic layers of alternating ferroelectric polarization having the director and polarization parallel to the layer planes, is exhibited by DIO [15]. The fourth new phase of this class of materials is the ferroelectric smectic A ($SmA_F$), also uniaxial with nearly saturated polarization [16,17].

We report here a transition from the ferroelectric nematic liquid crystal ($N_F$) phase to a low-temperature liquid phase with reentrant isotropic symmetry ($I_A$) in mixtures of the liquid crystal compound RM734 with few-percent concentrations of the ionic liquids BMIM-$PF_6$ (BMIM) or EMIM-TFSI (EMIM), shown in *Fig. 1*. Even a trace amount of ionic liquid dopant facilitates the kinetic pathway of the transition from $N_F$ to $I_A$, enabling simple cooling of the $N_F$ to produce the $I_A$ phase rather than resulting in crystallization. Unlike the $N_F$, the reentrant $I_A$ phase responds only weakly to applied electric field and thus appears to be nonpolar, *i.e.*, either paraelectric or antiferroelectric, with experiments to date suggesting the latter. The development of antiferroelectric ordering adjacent in temperature to the strongly polar-ordered $N_F$ phase may seem surprising, but the century-long experimental and theoretical study of ferroelectric materials, both crystals and liquid crystals, shows that ferroelectricity (F) and antiferroelectricity (A) go hand in hand, such that if one is found, the other will be observed in related materials, or even as coexisting phases. At the root of this behavior are dipole-dipole interactions and their inherent ambivalence: dipole pairs arranged end-to-end prefer to be oriented parallel, whereas dipole pairs arranged side-by-side prefer to be antiparallel. This frustration is almost a recipe for generating modulated, anisotropic, antiferroelectric phases of the SmZ$_A$ type [15], having layers of uniform *P* extended along the end-to-end direction and antiferroelectric ordering of adjacent layers. Here we probe the reentrance of isotropic symmetry from a highly polar and anisotropic N$_F$ liquid crystal state.

## *RESULTS*

The molecular species studied are shown in *Fig. 1A*. We carried out SAXS and WAXS experiments, depolarized transmission optical microscopy, and polarization current measurement of RM734/ionic liquid (IL) mixtures with the ionic liquid dilute, focusing here on RM734/BMIM at weight% ionic liquid concentrations, $c_{IL}$, in the range 0% ≤ $c_{IL}$ ≤ 10%. Unless stated otherwise, these experiments were carried out using a commercial RM734 sample (RM734$_\beta$) having a nematic (N) to ferroelectric nematic (N$_F$) transition temperature $T_{N\text{-}NF}$ = 128⁰C.

The phase diagram of *Fig. 1B*, constructed using x-ray and optical observations, summarizes our principal results on RM734$_\beta$/IL mixtures. We find that the introduction of modest amounts of ionic liquid can produce stunning changes in the phase behavior in the ferroelectric nematic realm. Even a very low concentration of ionic liquid ($c_{IL}$ ~ 0.01%) facilitates the appearance, upon lowering *T* below ~70⁰C, of a new bulk phase having reentrant isotropic symmetry, which we term the I$_A$, the "A" designating antiferroelectric symmetry, distinguishing this phase from the dielectric high-temperature isotropic phase (I) (*Fig. 1B*). Remarkably, even with such low concentrations of ionic liquid, the first-order thermotropic transition upon cooling from the N$_F$ phase to the I$_A$ phase fills the entire sample volume with I$_A$, suggesting that the I$_A$ phase may be intrinsic to the RM734$_\beta$ host, with the ionic liquid dopant serving to create a kinetic pathway for its nucleation. Additionally, at the phase boundary between the N and N$_F$ phases a additional thermotropic phase, a spatially modulated antiferroelectric (the M$_A$), appears, which, with increasing $c_{IL}$, supplants the ferroelectric nematic. This phase has same position in the phase diagram as the SmZ$_A$ phase has in DIO, and, like the SmZ$_A$ is layered and antiferroelectric with the director parallel to the layers, so it may be a SmZ$_A$ variant. The onset of phase separation of the ionic liquid component is observed for $c_{IL} \gtrsim 5\%$ (shaded region). Here we focus on the mesophase behavior at the lowest ionic liquid concentrations.

*X-ray scattering* – Powder-average SAXS and WAXS scans of temperature-controlled samples in 1 mm diameter, thin-wall capillaries were obtained upon slow cooling, with results shown in *Figs. 2, 6,* and *Figs. S1-S10* in the S*upplementary Information*. A selection of scans at temperatures in the N, M$_A$, N$_F$, and I$_A$, phases in the $c_{IL}$ = 1% mixture, as well as in the X phase of neat RM734, are shown in *Fig. 2A*. The complete set of X-ray scattering scans during cooling for all of the concentrations studied are available in the S*upplementary Information*. The gray-shaded peak at $q \sim 0.4$ Å$^{-1}$ is due to a Kapton window in the beam path, which is absent in room temperature scans (*Fig. 2B*). The higher temperature, anisotropic, N, M$_A$, and N$_F$ phases each exhibit a broad diffuse WAXS peak at $q \sim 1.5$ Å$^{-1}$, which arises from the powder average of the equatorial diffuse peak observable in the WAXS of magnetically-aligned N and N$_F$ RM734 samples reflecting side-by-side positional correlation due to steric repulsion of the molecules [3,9]. The transition to the I$_A$ phase results in significant changes in the scattering, including the appearance of a diffuse peak at small $q$ in the SAXS range (*Fig. 2B*, inset), and the breakup of the broad WAXS peak into several narrow, diffuse peaks, marking the development of specific, longer ranged, side-by-side intermolecular positional correlations in the I$_A$ phase. The small-angle peak is located at $q = q_M(T)$, where $q_M = 0.080$ Å$^{-1}$ at $T = 70^0$C, increasing to $q_M = 0.085$ Å$^{-1}$ at $T = 25^0$C. This peak position (corresponding to a length $2\pi/q_M = d_M \approx 75$ Å) is indicative of long-range, three-dimensional (3D) electron density modulation which, given the isotropy and transparency of the I$_A$ observed optically as discussed below, would be some form of short-ranged cubic lattice-like ordering with a unit cell of dimension comparable to $d_M$. However, this peak is diffuse, with a half-width at half maximum (HWHM) of $\delta q \sim 0.03$ Å$^{-1}$. The ratio $\delta q/q_M \sim 0.38$ is similar to that found in the static structure factor of hard spheres in 3D, generated by the interference of the scattering from each sphere with that of its shell of non-overlapping nearest neighbors [18]. In the present case the "spheres" would be unit-cell-like dynamic assemblies of tens to hundreds of molecules. The birefringent crystal (X) phase reappears only several days after returning the mixture to room temperature. These distinctive new features enable x-ray characterization of the I$_A$ and determination of its phase range.

The x-ray scattering $I(q)$ of the $I_A$ phase at $T = 25^0C$ in a $c_{IL}$ = 1% RM734$_\beta$ mixture is compared with that from samples with different ionic liquid concentrations in *Fig. 2B*. These scans show that the x-ray peak structure in the $I_A$ phase varies only very weakly with concentration for $c_{IL}$ in the range 0.2% ≤ $c_{IL}$ ≤ 10%, in both the SAXS and WAXS wavevector regimes. This, and the fact that the $I_A$ phase is observed at very low dopant concentrations, suggests that the causative molecular organization of the $I_A$ phase is intrinsic to the RM734$_\beta$ host.

*Optical textures and electro-optics of RM734 / BMIM mixtures* – Polarized light microscopy observations of the mixtures were made principally in glass cells with spacing $t$ in the range (3.5 μm < $t$ < 20 μm) between the glass plates. One plate was coated with a pair of ITO electrodes separated by a 1.04 mm-wide gap that were used to apply in-plane electric fields, $E$, largely parallel to the sample plane. The cell temperature $T$ was kept below 150$^0$C both during filling and afterwards, as higher temperatures damaged the components, which resulted in irreversible changes in phase behavior. The plates were treated with polyimide layers buffed antiparallel along a direction $3^0$ off from the electrode edges. This preparation produces uniform planar alignment of the N-phase director $n(r)$, the local mean molecular long axis and the optic axis, parallel to the glass and oriented $3^0$ from normal to the applied electric field. The $3^0$ offset ensures a well-defined initial field-induced reorientation direction of $n(r)$ with field application. Upon cooling into the $N_F$ phase, the $n(r)$-$P(r)$ couple adopts a π-twisted geometry, a result of the antipolar orientation of the ferroelectric polarization, $P(r)$, on the two surfaces imposed by the buffing [10]. Here we describe, by way of example, observations of two mixtures with different dopant concentrations in buffed cells and as a free drop on an untreated glass surface.

*Example 1: $c_{IL}$ = 0.2% mixture in a t = 3.5μm cell* – *Fig. 3* shows a twisted ferroelectric nematic monodomain in a $c_{IL}$ = 0.2% mixture being slowly cooled through the $N_F$ to $I_A$ phase transition at $T ≈ 66^0C$. The $I_A$ phase is dark between crossed polarizer and analyzer for all cell orientations, nucleating via a first-order transition as small domains, which eventually grow to cover the whole sample. The final $I_A$ state is dark when viewed between crossed polarizer and analyzer (*Fig 3D*) but exhibits a patchy texture of weakly transmitting areas of low remnant birefringence. Estimating this birefringence, which decreases in samples with increasing $c_{IL}$ and is undetectable when $c_{IL} > 1\%$, indicates it to be caused by sub-10 nm thick layers on the glass surfaces where there may be induced nematic order. In-plane applied electric fields of a few V/mm that reorient the director in the $N_F$ phase to be nearly along the field [11] have no visible effect on the dark domains.

*Example 2: $c_{IL}$ = 1% mixture in a t = 3.5µm cell* – The textures observed while cooling this mixture from the N to the $I_A$ phase, with and without applied electric field as shown in *Fig. 4*, exhibit the following features:

- *Paraelectric nematic phase (N)* – The uniformly aligned N monodomain of *Fig. 4A1* has a birefringence color in the first-order blue-to-green Michel-Levy band (retardance ~ 670 nm [19]), with the larger index for optical polarization along *n*. Application of an in-plane electric field of the order of 100 V/mm induces a twist Freedericksz transition [7] in the narrow LC isthmus between the two air bubbles (where *E* is largest), reorienting *n(r)* in the plane of the cell and reducing the birefringence color to a first-order orange-red (*Fig. 4A2*). The LC regions at either end of the isthmus are below the Freedericksz threshold field.

- *Antiferroelectric modulated phase ($M_A$)* – Upon cooling into the $M_A$ phase, the smooth N birefringence becomes textured by weak spatial non-uniformity, but with the average director still along the buffing (*Fig. 4A3* blue areas), forming "bookshelf" lamellar domains in a fashion similar to that of the $SmZ_A$ [15]. Now field-induced reorientation shows strong hysteresis as shown by the yellow area in *Fig. 4A3* at zero field, where previous application and removal of a large field reorganized this part of the texture into an array of permanently reoriented domains (*Fig. 4A3*). Subsequent application of the low fields with magnitudes comparable to those used to reorient the nematic, produce little additional change in the $M_A$ texture (*Fig. 4A4*), providing evidence that it is antiferroelectric. At lower temperatures in the $M_A$ phase, well-ordered stripe patterns with ~10 µm periodicity are observed (see *Figs. 5B,C*). These disappear at the transition to the $N_F$ phase.

- *Ferroelectric nematic phase ($N_F$)* – Upon cooling into the $N_F$ phase, the director-polarization field twists spontaneously between the top and bottom glass plates. The resulting texture is a mosaic of Grandjean-like, planar-aligned domains having the polarization parallel to the plates, but with varying magnitude and sign of twist (*Fig. 4A5*). These twist domains are extremely responsive to applied fields over the entire cell area (*Fig. 4A6*), even at the left- and right-hand ends of the LC isthmus where the applied field is smallest, of order 1 V/mm. This is typical ferroelectric nematic behavior in an antiparallel-buffed cell.

•*Antiferroelectric isotropic phase ($I_A$)* – The $N_F$ to $I_A$ phase transition is similar in appearance to that observed in the $c_{IL}$ = 0.2% cell described above, also being first-order and with cells exhibiting heterogeneous nucleation of the $I_A$ phase, which then grow as isolated isotropic domains (*Fig. 4B1*). With a cooling rate of -2$^0$C/min, these isotropic regions grows with time to cover the entire filled area of the cell in a few minutes. In contrast to the $c_{IL}$ = 0.2% cell described above, the degree of extinction of the $I_A$ region when viewed through crossed polarizers is comparable to that of the optically isotropic air bubbles, i.e., the $I_A$ regions have no detectable remnant birefringence. *Fig. 4B4* shows an image of the cell in transmission with the analyzer removed. The average brightness of the $I_A$ region is found to be 1.085 times that of the air bubbles. Given that the Fresnel reflections at the air/glass interfaces in the bubble area reduce the transmissivity of these regions by around 8%, this confirms that the intrinsic transmissivity of the $I_A$ phase is close to 1, i.e., that the $I_A$ is not strongly scattering.

<u>*Example 3: t ~ 30 µm thick free LC drop on an untreated glass substrate*</u> – A thick LC film enables the study of the light scattering characteristics of the $M_A$ phase. *Fig. 5A* shows a thick free drop on untreated glass plates being cooled through the N–$M_A$ and $M_A$–$N_F$ transitions viewed in transmission, with polarizer and analyzer as indicated. The N and $N_F$ phases (seen in *Figs. 5A1* and *A8,9* respectively) have defected textures but are only weakly scattering. The $M_A$ phase (seen in *Figs. 5A2* to *A8*), on the other hand, is initially strongly iridescent, scattering blue light which illuminates the cell and leaves the transmitted light primarily red, like the setting sun. This scattering is indicative of an internal modulated structure with an initial periodicity similar to that of the blue wavelength of light, and which shifts to longer wavelengths upon cooling toward the $N_F$, eventually passing through the visible range to the infrared (*Fig. 5A4*). Thus the modulation period $w_M(T)$ must increase upon cooling, shifting the scattered color toward longer wavelengths and toward smaller scattering angle. This concentrates the red scattering toward the forward direction, into the collection aperture of the microscope, as in *Figs. 5A5,5A6*. This increasing scale of the modulation period drives an increase in the absolute scattering cross-section, to such an extent that at sufficiently low temperature, the sample becomes opaque (*Fig. 5A7*). At the transition to the N$_F$, the sample becomes transparent again, as the scattering structure of M$_A$ phase transforms to the macroscopically defected texture of the N$_F$ by means of a first-order phase transition. At this transition, the modulation period grows to be much longer than visible wavelengths, the scattering regime is exited, and the visible transmission observed is the result of adiabatic propagation through a smoothly distorted medium in the N$_F$ phase (*Fig. 5A8,9*).

*Example 4: $c_{IL}$ = 1% mixture in an antiparallel-buffed t = 20 μm cell* – Filling the LC These observations suggest that the M$_A$ modulation period increases on cooling but provide no information on the nature of the modulated structure. However, the texture shown in *Fig. 5B*, with the mixture in a $t$ = 20 μm thick glass cell, provides strong evidence that the modulation is lamellar, with layers normal to the plates and on average aligned parallel to the buffing direction, *b*, like those of the SmZ$_A$ phase in DIO and its mixtures in similar cells [15]. This evidence is based on the behavior of thermotropic smectic lamellar phases which also have a modulation periodicity (smectic layer spacing) that is strongly temperature-dependent, such as the smectic C. When such lamellar phases are confined between flat plates with the layers normal to the plates (bookshelf alignment), it has been found that when the sample temperature is changed in such a way that the layers shrink (heating the M$_A$, in the present case), the layers will buckle in an effort to maintain the bookshelf pitch along the average layer normal direction [20], forming local chevron structures [21], and, on a larger scale, zig-zag walls [22]. The chevron texture relevant here is the case where the zig-zag walls run exactly normal to the layers [23], generating systems of internally zig-zag parallel stripes of dimension comparable to the sample thickness *t*. In the present case, upon heating the M$_A$ from the middle of its phase range to near the transition to the N, such undulations and long-range stripe patterns appear, as shown in *Fig. 5B,C*. The view in transmitted light without an analyzer (*Fig. 5B*) shows a texture of sharp undulations, the variations in color across the field of view suggesting the coexistence of several distinct layer spacings, as sketched in the inset. According to the model of [23], each such in-plane undulation of the layers generates a pair of stripes, each of width equal to *t*, which is approximately what is observed here. The stripes appear alternately red (zig)/green (zag) when viewed between crossed polarizer and analyzer (*Fig. 5C*). A small rotation of the sample causes the zig stripe color to interchange with the zag stripe color, providing additional visualization of the optic-axis orientation.

***Observation of the I$_A$ phase in undoped RM734*** – The observation that even very small concentrations of ionic liquid in RM734 suffice to facilitate the transition into the I$_A$ phase made us generally curious about the role of sample purity in determining the phase behavior of RM734. Indeed, the lack of substantial dependence of the WAXS scans on dopant concentration in ***Fig. 2*** strongly suggested that the I$_A$ structure could be a property of the RM734 host, and that neat RM734 could, in principle, exhibit the I$_A$ phase.

We therefore carried out experiments on three distinct batches of RM734 containing different impurities and having different degrees of impurity: $\alpha$, synthesized in our laboratory according to the scheme in Methods, having a nematic (N) to ferroelectric nematic (N$_F$) transition temperature $T_{N-NF}$ = 133⁰C, the temperature reported by Mandle et al. [3]; and two commercial samples, $\beta$ with $T_{N-NF}$ = 128⁰C; and $\gamma$ with $T_{N-NF}$ = 124⁰C. These transition temperatures suggest increasing impurity content in the $\alpha$, $\beta$, and $\gamma$ samples. Unless otherwise stated, all of the mixture data shown here and in the Supplementary Information were obtained using the RM734$_\beta$ sample.

We carried out experiments on all three batches in which the temperature vs. time cooling profiles were varied. Temperature profiles were found in each host which avoided crystallization on cooling to $T$ = 25⁰C and instead produced the I$_A$ phase (see ***Materials and Methods***). Crystallization was avoided with increasing ease the less pure the batch was, with RM734$_\gamma$ the least likely to crystallize. Samples of all three batches of RM734 eventually did crystallize but this process took days or weeks. The WAXS scans of the I$_A$ phase obtained in the three host batches at $T$ = 25⁰C are shown in ***Fig. 6***, along with the RM734$_\beta$/$c_{IL}$=1% BMIM mixture scan. The scans show only minor differences, supporting the notion that the I$_A$ phase is a property of RM734.

***Comparison of doping with BMIM and EMIM*** – Powder x-ray scans at $T$ = 25⁰C of RM734 samples mixed respectively with $c_{IL}$=5% BMIM and $c_{IL}$=5% EMIM are shown in ***Fig. 7***. The scans exhibit almost identical peak structures, providing further evidence that the I$_A$ phase is inherently a property of the host molecule. This result is consistent with the observed phase behavior and microscopic textures of the two mixtures, with EMIM doping producing essentially the same phenomenology in RM734 as BMIM doping.

*DISCUSSION*

*Reentrant isotropy* – The discovery of the $M_A$ and $I_A$ phases add an exciting dimension to the ferroelectric nematic realm. The $N_F$ phase, with its saturated quadrupolar order parameter and nearly perfect polar order even at elevated temperature, can be considered to be a well-ordered nematic phase. While the N, $M_A$, and $N_F$ phases combine long-range orientational order with short-range positional disorder, the $I_A$ phase lacks long-range orientational order but has exceptionally robust short-range molecular positional correlations (*Fig 2A*). The thermodynamic availability of the $I_A$ upon cooling points to internal energetic frustration in the $N_F$ as the principal driving force for this transition. This makes the $N_F - I_A$ phase transition a cousin of the nematic-to-helical antiferroelectric twist-bend (N-$TB_A$) transition, wherein a large internal energy cost in the packing of bent molecules in the N phase is traded for a more compact "brickwork" short-range packing motif in the $TB_A$ that has lower internal energy, a motif that stabilizes a different (helical) form of long-range ordering [24,25,26]. In the $N_F$-$I_A$ case, the change in the structure factor $I(q)$ upon passing from the $N_F$ to the $I_A$ is dramatic, as seen in *Fig 2A*, with the $I_A$ showing much more distinct, but not crystalline, diffuse peaks in the range (1 Å$^{-1}$ ≲ $q$ ≲ 2 Å$^{-1}$). Well-defined, diffuse peaks in this $q$-range are familiar in phases of rod-shaped or bent-core mesogens, where they arise from the side-by-side packing of molecules in smectic layers on short-range hexagonal [27] or herringbone lattices [28,29]. In the present case, a related energetically preferred local packing generates local correlations which stabilize long-range isotropy.

Thermotropic liquid crystal reentrant isotropy was first reported by Luzatti and Spegt in the cubic *Ia3d* gyroid phase of strontium alkanoates [30,31], a phase structure which was also found in the phenyl carboxylate dimers NO$_2$-*n*-BCA, upon cooling from a smectic C [32,33,34], and has been observed since then in numerous other systems [35,36,37,38]. The local gyroid structure is an assembly of molecules into extended supermolecular linear aggregates which form three interlinked sets of spring-like helices, one set running along each of the cartesian coordinate directions to generate a lattice of cubic symmetry made up of mesoscopic unit cells containing hundreds to thousands of molecules [30-38]. Their cubic symmetry renders such phases optically isotropic. The gyroid helices are chiral, which opens the possibility of isotropic phases having macroscopic chirality and consequent optical activity, even if the constituent molecules are achiral. Achiral molecules can also give rise to achiral isotropic phases if helical segments are found in mirror-symmetric pairs, as in the *Ia3d*.

Another specific example, which may be of particular relevance to the I$_A$ phase in RM734, is the fluid isotropic cubic phase obtained upon cooling the smectic C in the 1,2-bis( 4'*n*-alkoxybenzoyl) hydrazine (BABH-*n*) homologous series, also *Ia3d* [39,40,41,42,43,44,45,46,47]. In this system, at low temperatures the intermolecular hydrogen-bonding between side-by-side, rod-shaped cores enhances the hourglass shape of the molecules (thin cores with fat tails), and thereby the tendency for local twist, stabilizing a cubic phase having equivalent helical networks of opposite chirality [48]. Heating increases the side-by-side entropic repulsion of the flexible tails, breaking the intermolecular hydrogen bonds and producing a smectic C phase. Such effects of tail entropy can be enhanced by the use of multi-tail (phasmidic) molecules, making induction of local twist an effective method for obtaining cubic phases [36,37,49].

Extensive x-ray scattering on powder samples has been carried out on such cubic soft crystal systems in the low-Miller index range of their cubic lattice reciprocal spaces, generally with scattering vectors in the SAXS regime ($q \lesssim 0.4$ Å$^{-1}$). In many such cases, sets of scattering peaks can be indexed, and in some cases the peak intensities have been used to calculate unit cell electron density. The observed peak widths appear to be broader than the resolution limits, indicating that crystallite sizes, deduced from the inverse x-ray peak widths, are limited by lattice defects. However, optical observations show the growth of millimeter-dimension, cubic single crystals in some systems [36], providing direct evidence for long-range cubic crystal order.

In contrast, our SAXS scattering from the I$_A$ phase in the selection of RM734 samples, mixtures, and temperatures studied, exhibits a single peak at scattering vectors in the range (0.07 Å$^{-1}$ $\lesssim q_M(T) \lesssim 0.1$Å$^{-1}$), with a HWHM $\delta q \sim 0.03$ Å$^{-1}$, which is two orders of magnitude broader than the diffractometer resolution. Thus, this RM734 SAXS scattering peak is diffuse, indicating short-range positional correlations which, given this $q_M(T)$ range, have a length scale comparable to the unit cell sizes in the cubic systems cited above but without long-range cubic crystalline order. This situation is a good description of the "sponge phases" of lyotropic amphiphiles [50], and bent core smectics [51], and also observed in the Cubic*($Ia3d$) – Iso$_1$* – Iso$_2$ phase sequences of Tschierske and coworkers, observed in the family of achiral polycantenar mesogens which exhibit $Ia3d$ cubic phases with macroscopic spontaneous chirality [52,53]. The Iso$_1$* phase has a diffuse scattering peak at *q*-values where the cubic phases of this family have Bragg reflections, indicating short-range, cubic positional correlations in the Iso$_1$*. Remarkably, such correlations can maintain the macroscopic chirality of the cubic lattices, the Iso$_1$* exhibiting conglomerate domains of opposite optical rotation, indicating that they preserve and transmit their local chiral structure to their neighbors [52,53], in absence of a long-range lattice, as has also been found in bent cores [51]. The I$_A$ phase of RM734 may be an achiral example of such behavior, exhibiting similar cubic correlations, suggesting that a related, lower temperature phase with long-range cubic structure may exist.

While the SAXS data on cubic phases is extensive, there have been very few systematic studies of reentrant isotropics in the WAXS range ($q > 0.5$Å$^{-1}$), where larger scattering vectors can probe the details of molecular side-by-side packing, of relevance in our system. High-temperature phases such as the spontaneously chiral cubics have a broad, diffuse WAXS reflection [52], similar to that of the high-temperature N and N$_F$ phases of RM734.

However, we have found several experimental systems which exhibit low-temperature, isotropic polymorphism and have distinct, diffuse WAXS peaks very similar to those of the I$_A$ phase, the first in the 50 wt% mixture of the rod-shaped mesogen 8CB with the material W624 (compound **2b** in Ref. [54]), a thermotropic Smectic C – to – Isotropic* dimorphism was observed [55]. X-ray scattering and extensive freeze-fracture transmission electron microscopy visualization of the local structure in the spontaneously chiral Isotropic* (dark conglomerate) phase showed it to be locally lamellar with strong side-by-side molecular positional correlations and a strong tendency for local saddle-splay layer curvature, the latter driving the assembly of a disordered network of gyroid-like branched arrays of filaments of nested cylindrical layers [54], a form of sponge phase [50]. A similar low-*T* isotropic phase was also found in the neat samples of the bent-core molecule 12-OPIMB [55]. These phases feature director splay and saddle splay everywhere, which might work to stabilize such a phase in the ferroelectric nematic realm [56].

However, while these isotropics produce WAXS peaks similar to those of the $I_A$, their small-$q$ scattering is quite different from that of the $I_A$, their local nested-cylinder layer structure producing a sequence of smectic-like diffuse peaks in a 1:2:3 lamellar harmonic sequence [54], which is not observed in the RM734 $I_A$. A quite different low-$T$ isotropic system having x-ray structure more like the $I_A$ of RM734 is described in Ref. [57], which reports a small azo-based molecule (W470), which forms side-by-side linear aggregates and gels when diluted with solvent. However, the structure of the reentrant isotropic phase of W470 has not been established.

The soft cubic phases discussed above are based on nanosegregation, which in the gyroid case involves the formation of filamentous networks. Other nanosegregation motifs which may be relevant to understanding the $I_A$ phase are the formation of localized aggregates, such as micelles or localized topological singularities, which then self-assemble into higher-symmetry arrays [30]. Experimental demonstrations include crystals of complex defects such as skyrmions [58] and knots [59], block-copolymers [60], and hybrid thermotropic 3D variations, the so-called "transparent nematic" in which dispersed didodecylammonium bromide micelles serve as the cores of fluid hedgehog defects in a rod-shaped mesogenic nematic [61,62]. Periodic arrays of nematic topological defects have been shown theoretically to form stable, space-filling crystal structures [56,63,64,65,66].

<u>$M_A$ phase</u> –While the $I_A$ phase shows distinctive microscopic textures and characteristic x-ray scattering features, the WAXS powder structure function of the $M_A$ phase is very similar to those of the N and $N_F$ phases, which bound it in the phase diagram, so that the $M_A$ is difficult to distinguish by x-ray scattering. The $M_A$ phase is observable in the microscope at higher ionic dopant concentration (for example, for $c_{IL} \geq 0.2\%$, as in *Fig. 5*) but is not distinguishable optically at smaller concentrations or in undoped RM734. As is evident from *Fig. 1*, the $M_A$ phase temperature range is roughly proportional to the ionic liquid concentration, making it an equilibrium phase of the mixtures.

This behavior is similar to that of the $SmZ_A$ phase in RM734/DIO mixtures, a phase that is also intermediate between the N and $N_F$, and is also spatially modulated. The published analyses of the $SmZ_A$ phase of DIO [10] and its mixtures with RM734 [9,67,68] may therefore offer insight into the appearance of the M$_A$ in the RM734/ionic liquid mixtures. While the initial polarized microscopy study of the SmZ$_A$ phase in RM734/DIO mixtures showed the SmZ$_A$ persisting only down to ~50 wt% DIO [9], subsequent precision adiabatic calorimetry showed that this intermediate phase between the N and N$_F$ could be detected at DIO concentrations as low as 10 wt% [67,68]. We have only begun to study the structural properties of the M$_A$ phase, but what is known from the discussion above and *Fig. 5* is that it shares its principal characteristics with the SmZ$_A$: •found between the N and N$_F$ phases; •lamellar; •antiferroelectric; •in cells with rubbed alignment surfaces, the M$_A$ phase forms "bookshelf" lamellar domains with the director along the rubbing and the layers normal to the cell plates, as shown in *Figs. 5B,C*, in a fashion identical to that of the SmZ$_A$ [15]. Cooling of the M$_A$ phase produces layer contraction, which in turn produces layer undulations and a chevron texture [23], exhibiting direct evidence of layering, like the SmZ$_A$. These similarities lead us to tentatively identify the M$_A$ as a SmZ$_A$ phase, or some SmZ$_A$ variant having a large modulation period with strong temperature dependence.

Recalling the Landau model discussed in Ref. [11], which successfully characterizes the overall features of the N – SmZ$_A$ – N$_F$ temperature/electric field phase diagram in DIO, predicts as a basic requirement that, as the terminal concentration of the SmZ$_A$ phase range (the Lifshitz point) is approached with increasing RM734 concentration, the wavelength of the SmZ$_A$ modulation will diverge. However, this same model predicts only a weak temperature dependence of the modulation period of the SmZ$_A$ phase, as found in DIO [15], which is in contrast to the optical observations of the RM734/BMIM mixture shown in *Fig. 5*.

## MATERIALS AND METHODS

The mixtures were studied using standard liquid crystal phase analysis techniques previously described [7,10,15,16], including polarized light microscope observation of LC textures and their response to electric field, x-ray scattering (SAXS and WAXS), and techniques for measuring polarization and determining electro-optic response.

*Materials* – Samples of RM734 (4-[(4-nitrophenoxy)carbonyl]phenyl 2,4-dimethoxybenzoate), shown in *Fig. 1* and first reported in Ref. [6], from three independent sources were tested in this study. $RM734_\alpha$ was synthesized by the Walba group as described in [9] ($RM734_\alpha$ phase sequence: I – $188^0$C – N – $133^0$C – $N_F$ – ≲$100^0$C – X); $RM734_\beta$, from Instec, Inc., was used as received ($RM734_\beta$ phase sequence: $179^0$C – N – $128^0$C – $N_F$ – ≲$100^0$C – X); $RM734_\gamma$, a third preparation also from Instec, Inc., was used as received ($RM734_\gamma$ phase sequence: I – $177^0$C – N – $124^0$C – $N_F$ – ≲$100^0$C – X). These transition temperatures suggest that the $\alpha$, $\beta$, and $\gamma$ samples have decreasing purity. This ranking is also seen in decreasing $N_F$ – X transition temperatures, although these varied widely in each sample depending on sample volume and cooling rate, as is typical LC crystallization behavior. BMIM-PF6 (1-Butyl-3-methylimidazolium hexafluorophosphate), and EMIM-TFSI (1-Ethyl-3-methylimidazolium bis(trifluoromethylsulfonyl)imide) were obtained from Millipore/Sigma and used without further purification.

*Methods* – RM734 / IL mixtures for cells and capillaries were prepared in the N phase but were never heated above $T$ = 150°C as higher temperatures damaged the components, leading to irreversible changes in phase behavior. X-ray and DTOM temperature scans were carried out on cooling. The capillaries were cooled to room temperature to record the diffraction patterns, and then heated up again to the paraelectric nematic phase and cooled at -0.5°C per minute for additional measurements at selected temperatures until the $I_A$ phase was reached, where they were held at constant temperature for a period of duration sufficient to allow droplets of the $I_A$ phase to coalesce and anneal.

*X-ray scattering* – For SAXS and WAXS, the LC samples were filled into 1 mm-diameter, thin-wall capillaries. The scans presented here are powder averages of diffractograms obtained using a Forvis microfocus SAXS/WAXS system with a photon energy of CuK$_\alpha$ 8.04 keV (wavelength = 1.54 Å).

*Depolarized transmission optical microscopy (DTOM)* – DTOM of LC cells in transmission between crossed polarizer and analyzer, with such cells having the LC between uniformly spaced, surface-treated glass plates, enables direct visualization of the director field, *n(r)*, and, apart from its sign, of *P(r)*. The observed uniaxial optical textures and their response to electric fields provide key evidence for the macroscopic polar ordering and fluid layer structure of LC phases.

*Electro-optics* – For making electro-optical measurements, the mixtures were filled into planar-aligned, in-plane switching test cells (Instec, Inc.) with unidirectionally buffed alignment layers arranged antiparallel on the two plates, which were uniformly separated by *t* in the range 3.5 μm < *t* < 8 μm. In-plane ITO electrodes on one of the plates were spaced by a 1 mm wide gap and the buffing was along a direction rotated 3° from parallel to the electrode edges. Such surfaces give a quadrupolar alignment of the N and SmZ$_A$ directors along the buffing axis and polar alignment of the N$_F$ at each plate. The antiparallel buffing makes antipolar cells in the N$_F$ phase, generating a director/polarization field that is everywhere parallel to the plates but undergoes a π-twist between the plates [10].

*ACKNOWLEDGEMENTS*

This work was supported by NSF Condensed Matter Physics Grants DMR 1710711 and DMR 2005170, by Materials Research Science and Engineering Center (MRSEC) Grant DMR 1420736, by the State of Colorado OEDIT Grant APP-354288, and by a grant from Polaris Electro-Optics. X-ray experiments were performed in the Materials Research X-Ray Diffraction Facility at the University of Colorado Boulder (RRID: SCR_019304), with instrumentation supported by NSF MRSEC grant DMR-1420736.

FIGURES

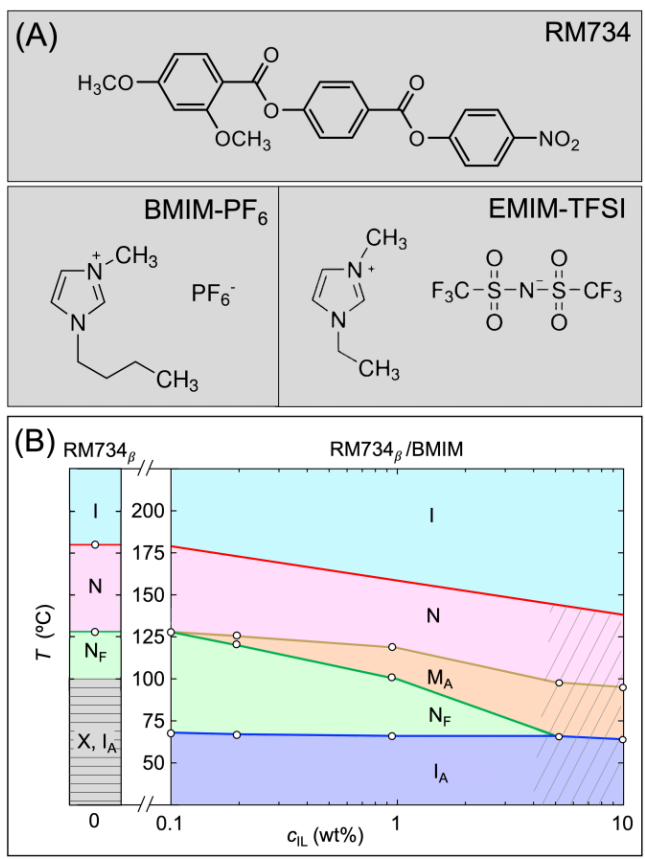

*Figure 1*: (*A*) Structures of the chemical species studied. (*B*) The phase diagram of RM734 and RM734/BMIM mixtures, including the $I_A$ and $M_A$, new phases of the ferroelectric nematic realm. The molecule RM734 is mesogenic, exhibiting the following phase sequence on cooling from 200ºC: high-temperature dielectric isotropic (I); paraelectric nematic (N); ferroelectric nematic ($N_F$); and, depending on the cooling process, crystal (X) or low-temperature, antiferroelectric isotropic ($I_A$). The phase diagram was obtained from SAXS and WAXS experiments and optical microscopy observations of RM734$_\beta$/BMIM mixtures (see text). Exposing the mixtures to temperatures higher than T ~ 150ºC produces irreversible changes in phase behavior, so the I – N transition temperature in the mixtures was evaluated only approximately. The $I_A$ phase is liquid when it first appears on cooling, becoming gel-like and then glassy as T is decreased. The times and temperatures required for crystallization were highly dependent on the ionic liquid concentration $c_{IL}$ and on sample geometry, with longer times and lower temperatures required at higher dopant concentrations. The mixtures exhibit a lamellar, modulated, antiferroelectric phase ($M_A$) not observed in neat RM734. At sufficiently high BMIM concentrations (shaded region), phase separation of the dopant material is observed.

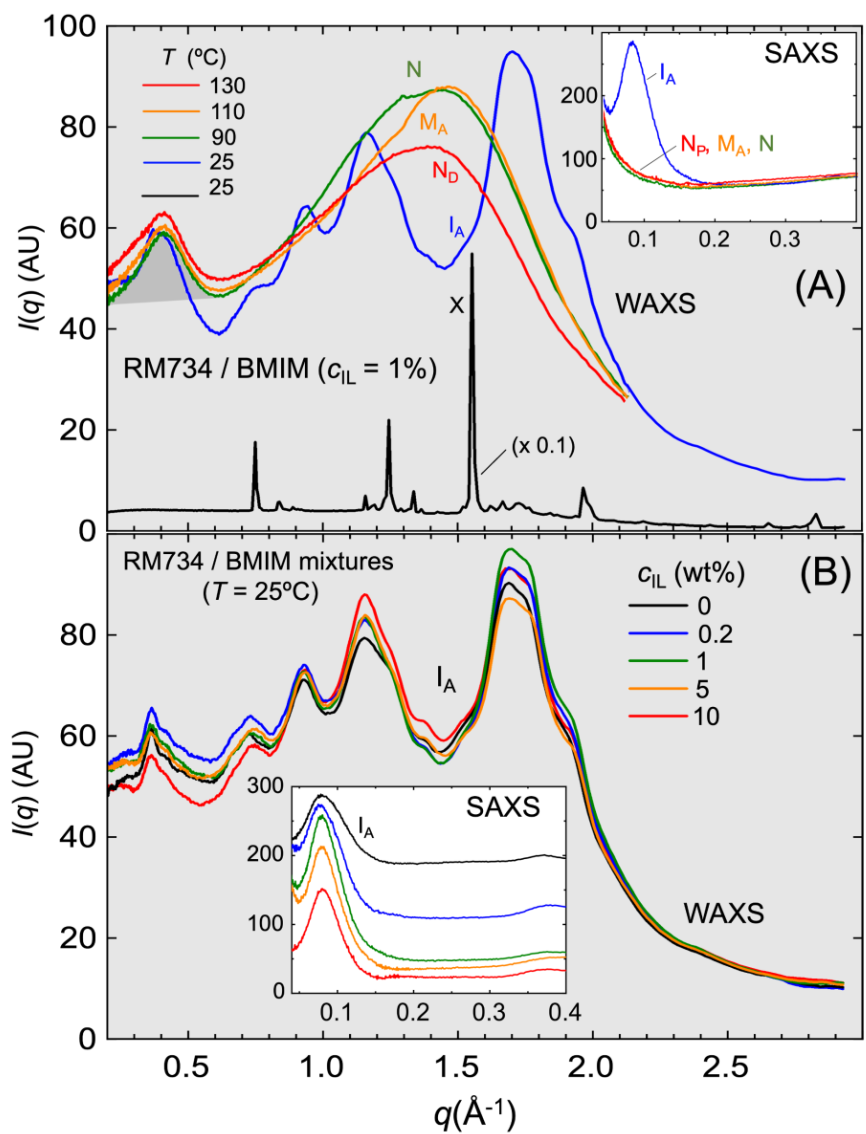

*Figure 2*: SAXS and WAXS scans obtained during slow cooling of RM734/ BMIM mixtures. The $I_A$ phase exhibits a diffuse peak at small wavector ($q \sim 0.07$ Å$^{-1}$, SAXS insets) indicating supermolecular periodicity ($d_M \sim 80$ Å), and a distinctive pattern of diffuse peaks in the $q$-range of side-by-side molecular packing but with wavevectors very different from those of the crystal (X) phase. (*A*) Scans obtained on a $c_{IL} = 1$ wt% RM734/ BMIM mixture at selected $T$ values show the observed structure factor $I(q)$ typical of each of the N, $M_A$, $N_F$, $I_A$, and X phases. The complete set of scans vs. $T$ is shown in the Supplementary Information. The shaded peak at $q \sim 0.4$ Å$^{-1}$ is due to a Kapton window in the x-ray beam path. The scans show significant differences in the x-ray structure function of the $I_A$ from that of the X and the higher $T$ phases. (**B**) Scans obtained at $T =$ 25 °C for selected $c_{IL}$ values show that the $I_A$ phase structure function depends only weakly on BMIM concentration. The SAXS scan baselines are shifted for clarity.

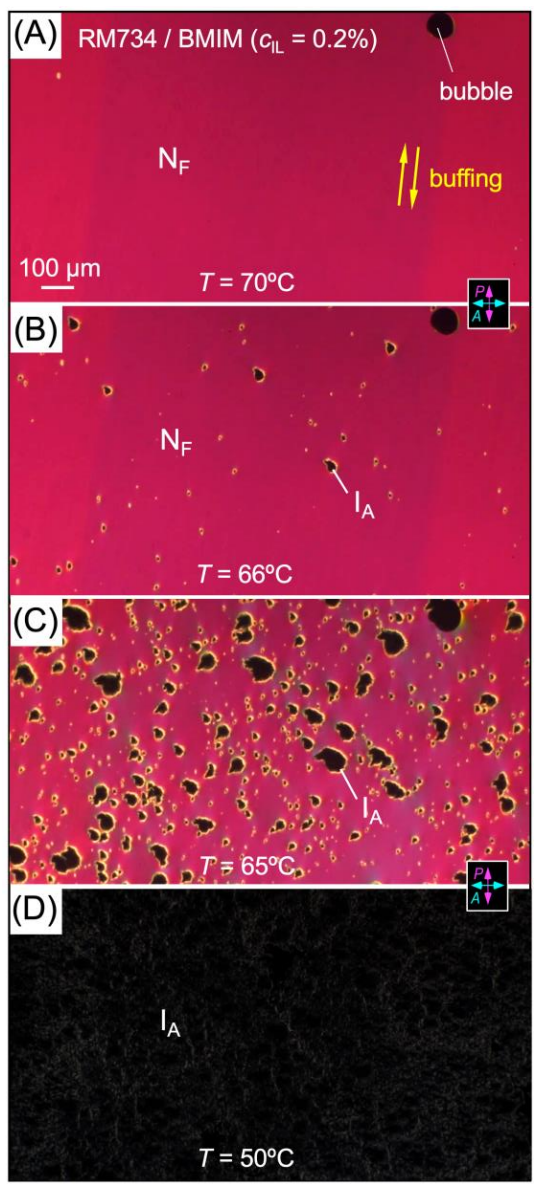

*Figure 3*: Depolarized transmission optical microscopy images of a 3.5 μm thick RM734$_\beta$/$c_{IL}$ = 0.2% BMIM mixture between glass plates with antiparallel rubbed alignment layers. (**A**) In the N_F phase, this surface treatment results in a uniform π-twist of the director from one plate to the other, producing a pink birefringence color between crossed polarizer and analyzer. (*B,C*) Upon continuous, slow cooling from the N_F phase, at temperatures below ~70⁰C dark domains of the I_A phase begin to appear, eventually covering the entire cell area (*D*). In such low-concentration mixtures, a thin birefringent layer a few nanometers thick remains on the sample surfaces when the bulk is in the I_A phase. This layer is not observed for higher BMIM concentrations, as in *Figure 4*.

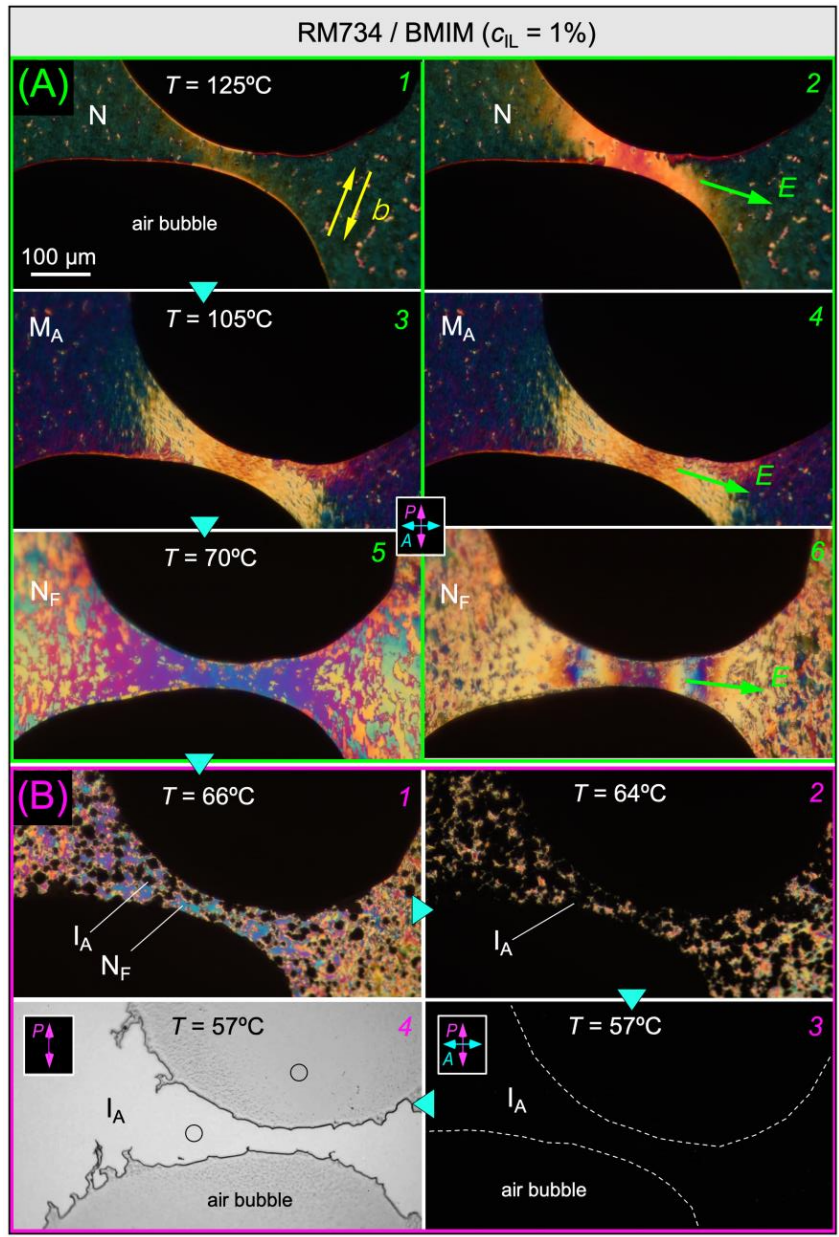

*Figure 4*: Depolarized transmitted light image of a 3.5 μm thick RM734 sample between glass plates with antiparallel-rubbed alignment layers. This surface treatment produces nearly uniform director alignment in the N and M$_A$ phases, and a director structure that is uniformly π-twisted from one plate to the other [10] in the N$_F$. A pair of rectangular evaporated electrodes on one cell plate are spaced by 1 mm, enabling application of an in-plane electric field *E*.

(*A*) Effect of applied field in the N, M$_A$, and N$_F$ phases. (A1 → A2) *Paraelectric nematic phase (N)*: Applied field induces twist of *n* in the interior of the cell; the twist relaxes away once the field is removed. (A3) *Modulated antiferroelectric phase (M$_A$)*: Applied field has induced twist of *n* in the interior of the cell, which does not relax away once the field is removed, evidence that the M$_A$ is structured. (A3 → A4) *Modulated antiferroelectric phase (M$_A$)*: Following initial alignment of *n* with a large in-plane field as indicated, there is little response to fields applied subsequently with strengths comparable to those that readily reorient the N phase, evidence that the M$_A$ is antiferroelectric; (A5 → A6) *Ferroelectric nematic phase (N$_F$)*: Small applied fields alter the planar twist of *n* everywhere in the sample; the twist sense may be reversed by changing the sign of the field. (*B*) Nucleation and growth of the I$_A$ phase. (B1 → B3) At low temperatures, the antiferroelectric isotropic (I$_A$) phase grows into the N$_F$. At this relatively large ionic liquid concentration, there is no observable remnant birefringence in the I$_A$ phase, even in applied fields of up to 100 V/mm. (B4) The I$_A$ phase, viewed here without an analyzer, is essentially transparent, transmitting 8.05% more light than the surrounding air bubbles. This is consistent with the I$_A$ showing very little scattering and nearly refractive index-matching the glass plates, while the intensity of light passing through the bubbles is reduced by two Fresnel reflections at the air/glass interfaces.

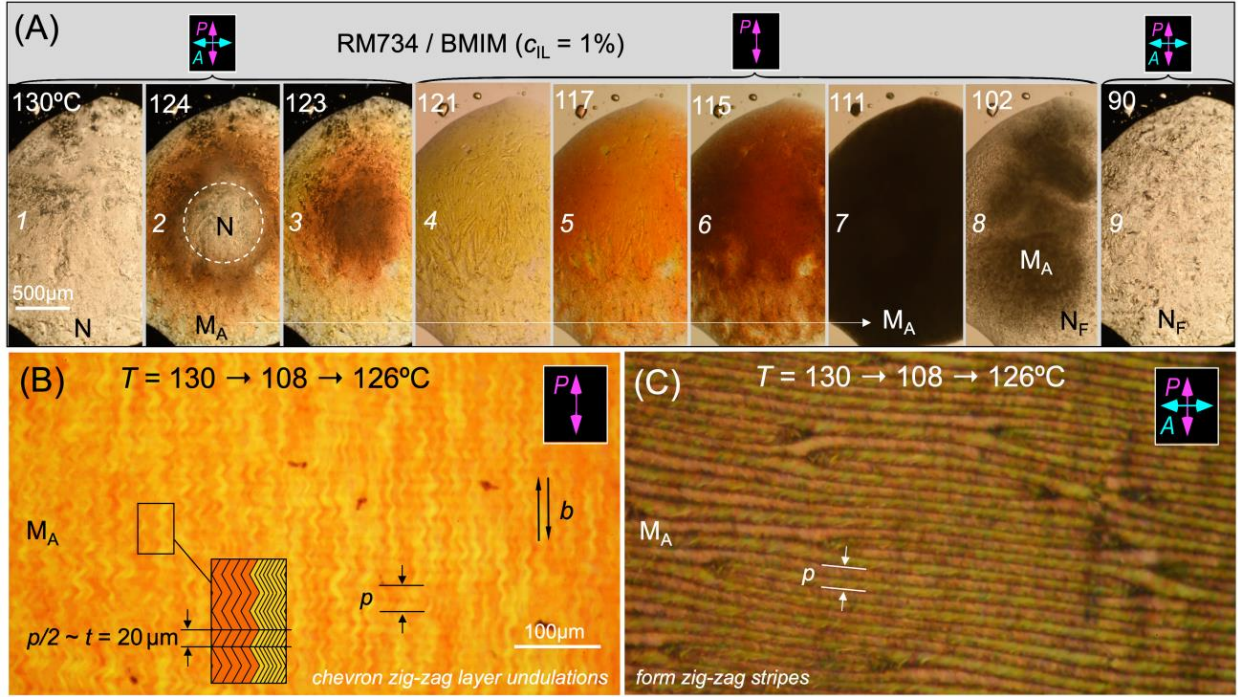

*Figure 5*: Transmission optical microscope images of $c_{IL}$ = 1% RM734/BMIM mixtures. (*A*) Optical appearance of a $t \sim 30$ μm thick free drop on an untreated glass substrate on cooling. (*A1*) Disordered nematic texture. (*A2-3*) Red light transmission at high $T$ in the $M_A$ phase, due to scattering of short wavelength light; (*A4-6*) Visible light scattering moving to larger wavelength and smaller scattering angle with decreasing $T$. (*A7*) Strong scattering makes the sample opaque; (*A8*) Coarsening domains become visible in the microscope; (*A9*) Disordered ferroelectric nematic texture. (*B,C*). Layer undulations and modulated textures in the $M_A$ phase between buffed glass plates (antiparallel buffing *b*) spaced by $t = 20$ μm. Cooling the sample into the $M_A$ phase and then heating it from the middle of the $M_A$ phase range to near the transition to the N and holding it causes stable patterns of layer undulations and extended stripes to appear. The texture in transmitted light without an analyzer is a mosaic of red-orange and yellow sawtooth-like undulations, suggesting the development of several distinct, coexisting layer spacings which scatter different wavelengths of light, as suggested schematically in the inset in (*B*), under the dynamic conditions of forced layer shrinkage. In the chevron model response to layer shrinkage, developed to describe smectic C textures [21,22,23], such zig-zag defects mediate the formation of stripes, each of width comparable to the sample thickness, *t*, which is approximately what is observed here.

-22-

(C) shows that the zig-zag texture within the stripes can be observed with crossed polarizer and analyzer. The red/green colors of the stripes interchange when the sample is rotated by a few degrees, providing additional visualization of the zig-zag optic-axis orientation.

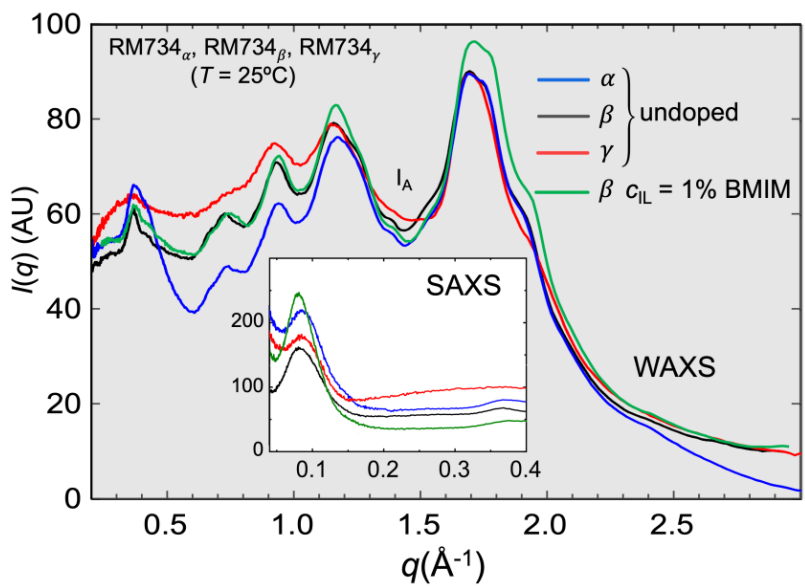

*Figure 6*: Scans showing the WAXS x-ray structure functions $I(q)$ of the low-temperature isotropic ($I_A$) phase at $T = 25^0C$ in three neat (i.e., undoped) RM734 samples, along with the $T = 25^0C$ scan of the RM734$_\beta$ sample doped with $c_{IL}$ = 1wt% BMIM. The intensities are scaled such that the heights of the $q = 1.7$Å$^{-1}$ peaks match. The RM734$_\alpha$ scan has a Kapton peak at $q = 0.4$ Å$^{-1}$ which is absent in the other measurements. The scattering patterns of all four samples are similar, implying that the three neat samples of RM734 have $I_A$ phase structures very similar to that of the doped liquid crystal.

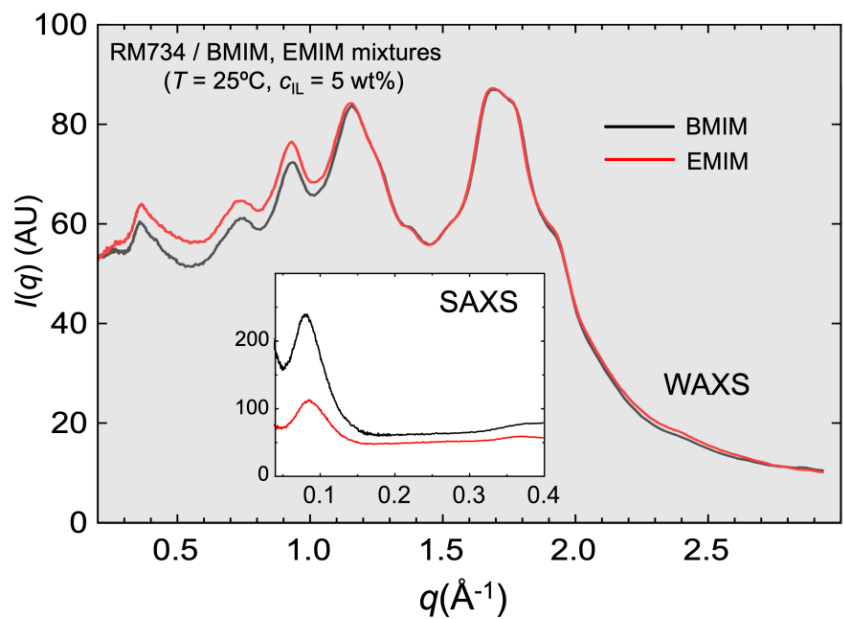

*Figure 7*: X-ray scattering from RM734 doped with two different ionic liquids. These scans show the WAXS x-ray structure functions at $T = 25°C$ of the $I_A$ phase of RM734$_β$ samples doped respectively with $c_{IL}$ = 5 wt% BMIM and $c_{IL}$ = 5 wt% EMIM, scaled such that the $q = 1.7 Å^{-1}$ peaks have the same height. The two samples produce very similar scattering, indicating that BMIM and EMIM facilitate formation of $I_A$ phases with very similar structure in RM734.

*Thermotropic reentrant isotropy and antiferroelectricity in the ferroelectric nematic material RM734*

Xi Chen[1], Min Shuai[1], Bingchen Zhong[1], Vikina Martinez[1], Eva Korblova[2], Matthew A. Glaser[1], Joseph E. Maclennan[1], David M. Walba[2], Noel A. Clark[1]*

[1]*Department of Physics and Soft Materials Research Center,*
*University of Colorado, Boulder, CO 80309, USA*

[2]*Department of Chemistry and Soft Materials Research Center,*
*University of Colorado, Boulder, CO 80309, USA*

*Abstract*

We report a transition from the ferroelectric nematic liquid crystal ($N_F$) phase to a lower-temperature, antiferroelectric fluid phase having reentrant isotropic symmetry ($I_A$), in the liquid crystal compound RM734 doped with small concentrations of the ionic liquids BMIM or EMIM. Even a trace amount of ionic liquid dopant facilitates the kinetic pathway for the transition from the $N_F$ to the $I_A$, enabling simple cooling to produce this isotropic fluid phase rather than resulting in crystallization. The $I_A$ was also obtained in the absence of specific ionic liquid doping by appropriate temperature cycling in three distinct, as-synthesized-and-purified batches of RM734, two commercial and one from our laboratory. An additional birefringent, lamellar-modulated, antiferroelectric phase with the director parallel to the layers, resembling the smectic $Z_A$, is found between the paraelectric and ferroelectric nematic phases in RM734/BMIM mixtures.

## TABLE OF CONTENTS

The mixtures were studied using standard liquid crystal phase analysis techniques, previously described [1,2,3,4,5] including Depolarized Transmission Optical Microscopic (DTOM) observation of LC textures and response to electric field, x-ray scattering (SAXS and WAXS), and techniques for measuring polarization and determining electro-optic response [4].

*Materials* – Samples of RM734 (4-[(4-nitrophenoxy)carbonyl]phenyl 2,4-dimethoxybenzoate), shown in *Fig. 1* and first reported in Ref. [6], from three independent sources were tested in this study. RM734$_\alpha$ was synthesized by the Walba group as described in [2] (RM734$_\alpha$ phase sequence: I – 188$^0$C – N – 133$^0$C – N$_F$ – ≲100$^0$C – X); RM734$_\beta$, from Instec, Inc., was used as received (RM734$_\beta$ phase sequence: 179$^0$C – N – 128$^0$C – N$_F$ – ≲100$^0$C – X); RM734$_\gamma$, a third preparation also from Instec, Inc., was used as received (RM734$_\gamma$ phase sequence: I – 177$^0$C – N – 124$^0$C – N$_F$ – ≲100ºC – X). These transition temperatures suggest that the $\alpha$, $\beta$, and $\gamma$ samples have decreasing purity. This ranking is also seen in decreasing N$_F$ – X transition temperatures, although these varied widely in each sample depending on sample volume and cooling rate, as is typical LC crystallization behavior. BMIM-PF6 (1-Butyl-3-methylimidazolium hexafluorophosphate), and EMIM-TFSI (1-Ethyl-3-methylimidazolium bis(trifluoromethylsulfonyl)imide) were obtained from Millipore/Sigma and used without further purification.

*Methods* – RM734 and its mixtures with ionic liquid were heated to 150°C before they were loaded into capillaries or liquid crystal cells. Higher temperatures damaged the components producing irreversible changes in phase behavior. The samples were then heated on a heating stage to the uniaxial nematic phase at 150 to 120°C and subsequently cooled down to 70°C, around which temperature the dark phase starts to grow. For X-ray diffraction experiments, the capillaries were kept at 65°C for about 10 minutes to allow the dark phase to anneal. The capillaries were cooled to room temperate to record the diffraction patterns, and then heat up again to uniaxial nematic phase and cooled at 0.5°C per minute to various temperatures for data recording.

*X-ray scattering* – For SAXS and WAXS LC samples were filled into 1 mm diameter, thin-wall capillaries. The data presented here are powder averages obtained using a Forvis microfocus SAXS/WAXS system with a photon energy of CuK$_\alpha$ 8.04 keV (wavelength = 1.54 Å).

*Depolarized transmission optical microscopy (DTOM)* – Observation of LC cells in transmission between crossed polarizer and analyzer, with such cells having the LC between uniformly spaced, surface treated glass plates, provides key evidence for the macroscopic polar ordering, uniaxial optical textures, and fluid layer structure in LC phases. DTOM enables direct visualization of the director field, *n*(*r*), and, apart from its sign, of *P*(*r*).

*Electro-optics* – For making electro-optical measurements, the mixtures were filled into planar-aligned, in-plane switching test cells with unidirectionally buffed alignment layers arranged antiparallel on the two plates, which were uniformly separated by *d* in the range 3.5 μm < *d* < 20 μm. In-plane ITO electrodes were spaced by a 1 mm wide gap and the buffing was parallel to the gap. Such surfaces give a quadrupolar alignment of the N and SmZ$_A$ directors along the buffing axis and polar alignment of the N$_F$ at each plate. The antiparallel buffing makes *antipolar* cells in the N$_F$, generating a director/polarization field parallel to the plates, and with a π-twist between the plates [3].

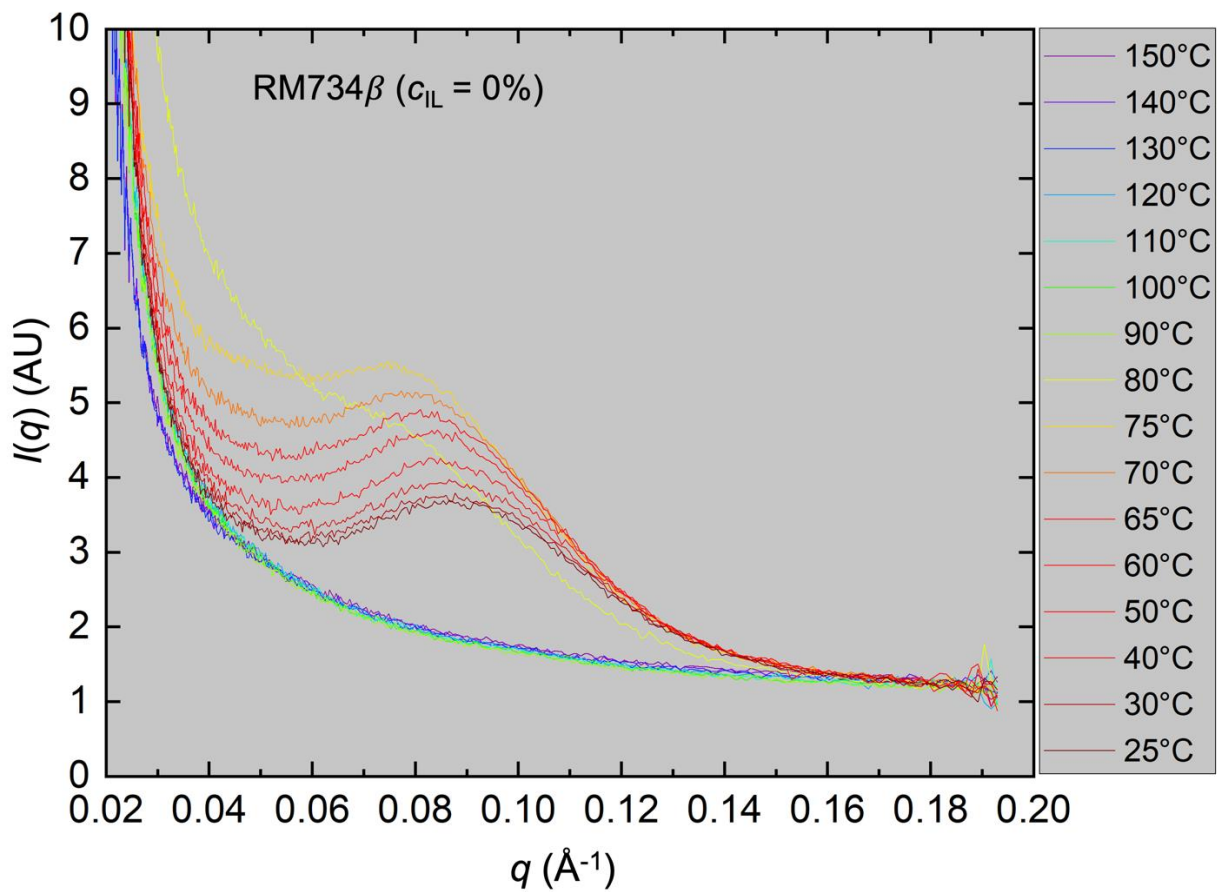

*Figure S1:* SAXS scans vs. *T* of undoped RM734$_\beta$.

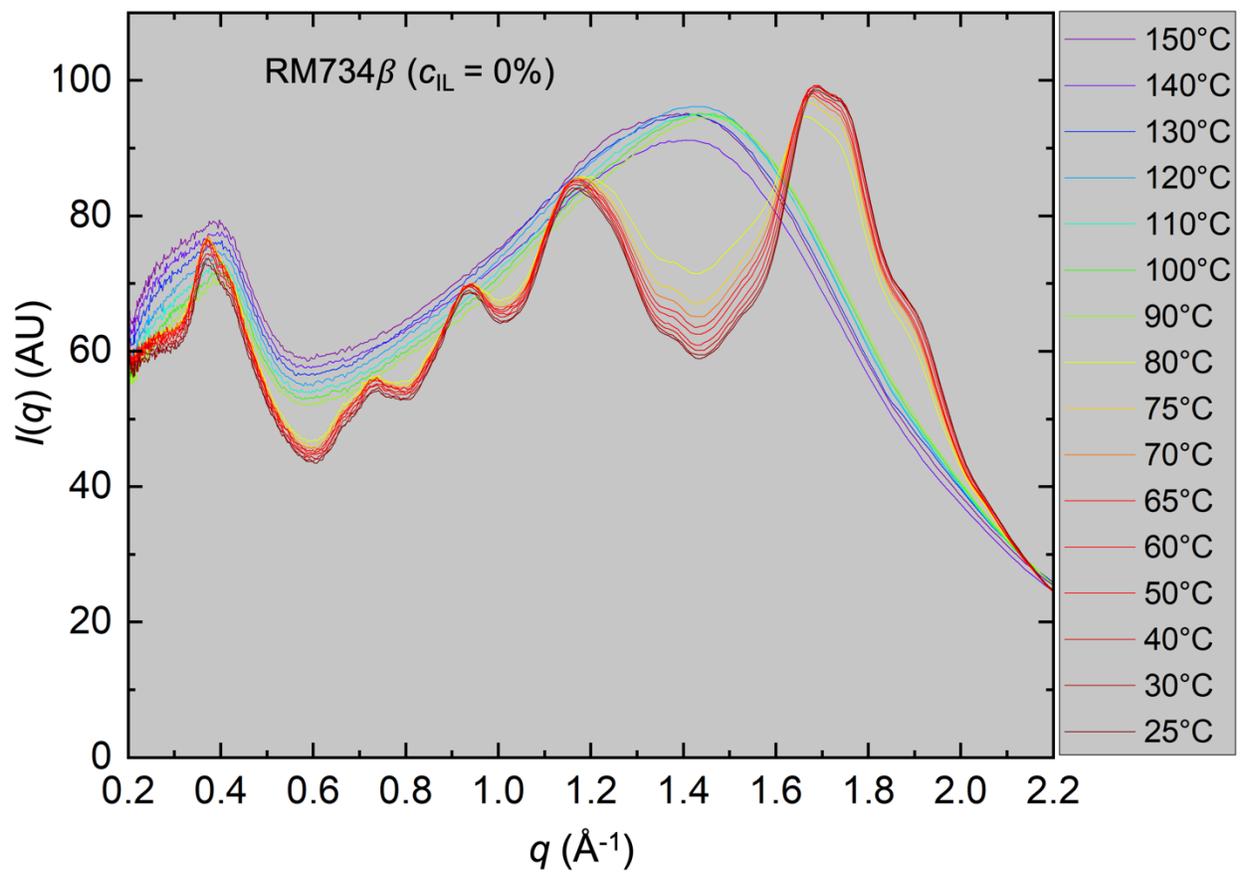

*Figure S2:* WAXS scans vs. $T$ of undoped RM734$_\beta$.

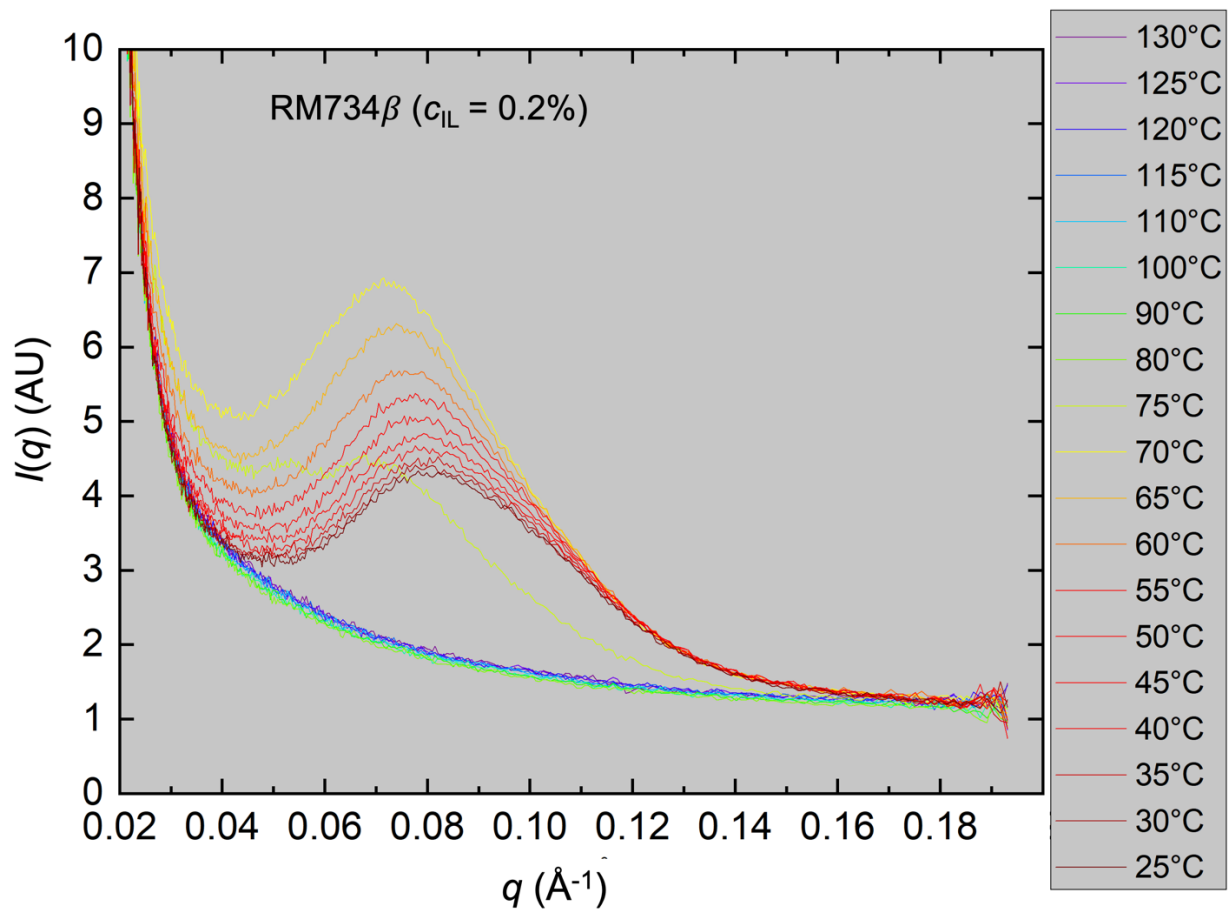

*Figure S3:* SAXS scans of RM734$_\beta$/$c_{IL}$ = 0.2 wt% mixture.

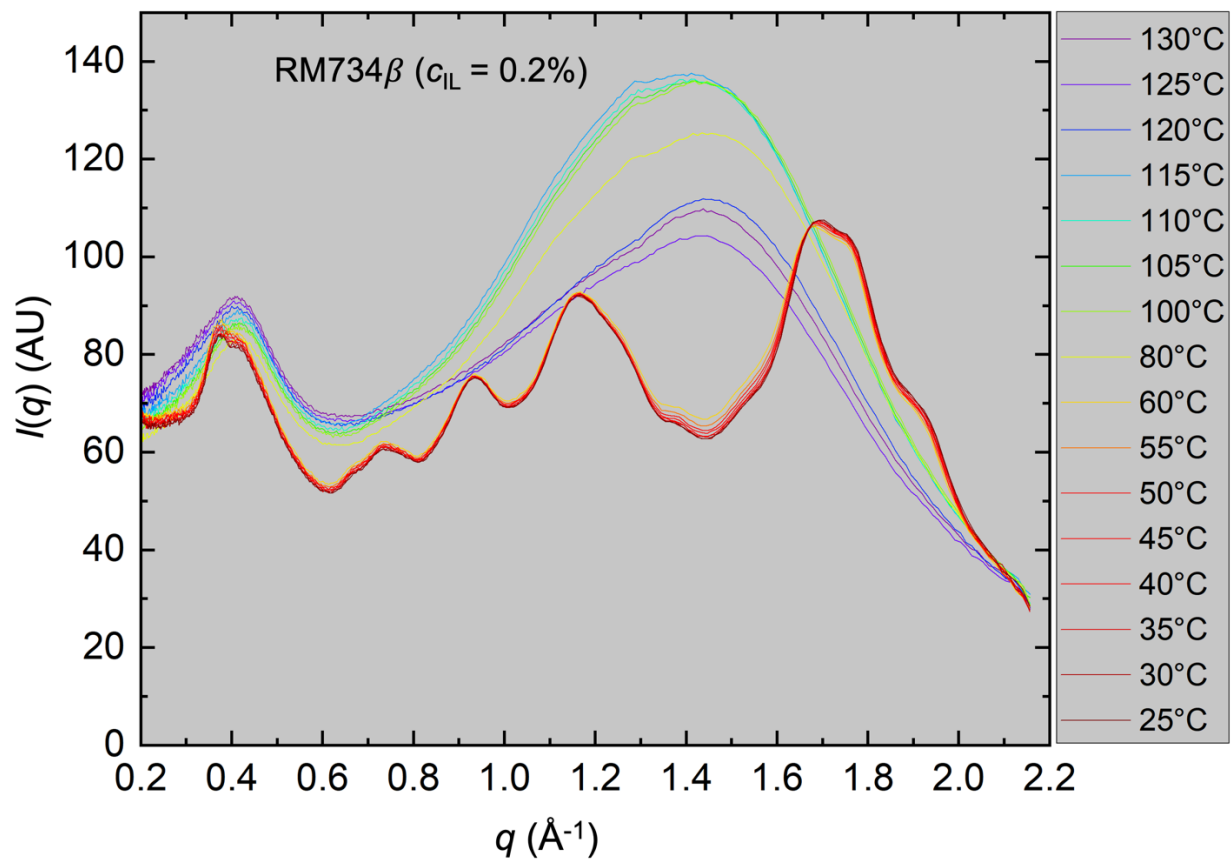

*Figure S4:* WAXS scans of RM734$_\beta$/$c_{IL}$ = 0.2 wt% mixture.

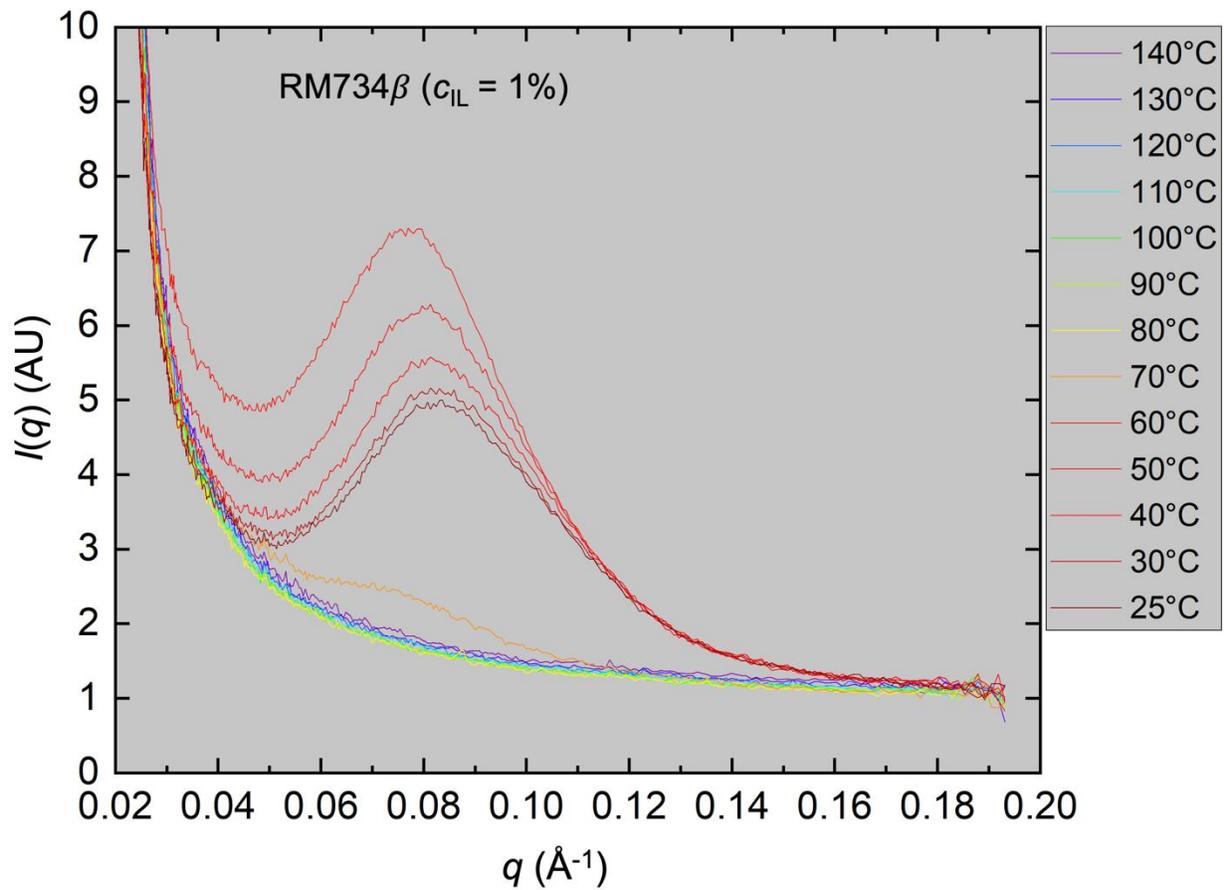

*Figure S5:* SAXS scans of RM734$_\beta$/$c_{IL}$ = 1 wt% mixture.

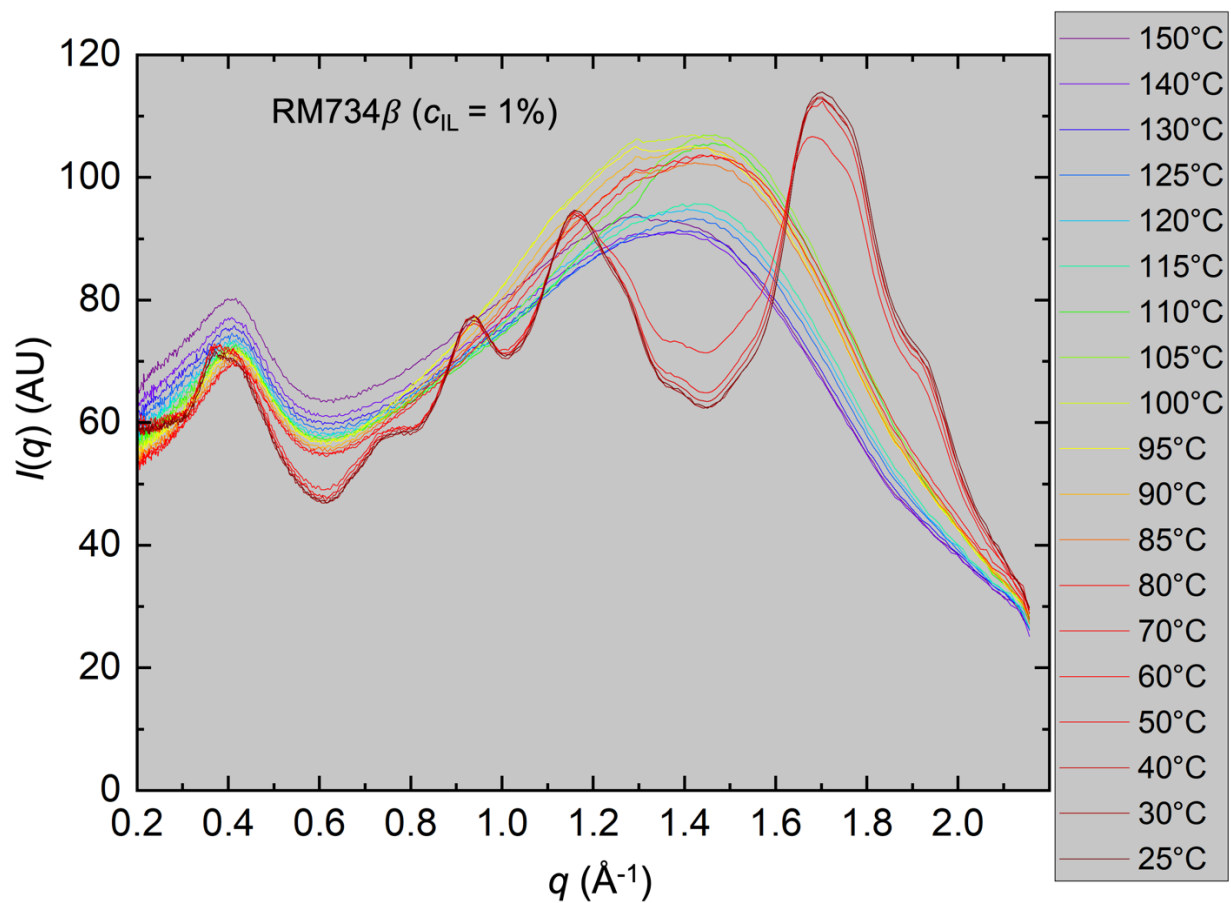

*Figure S6:* WAXS scans of RM734$_\beta$/$c_{IL}$ = 1 wt% mixture.

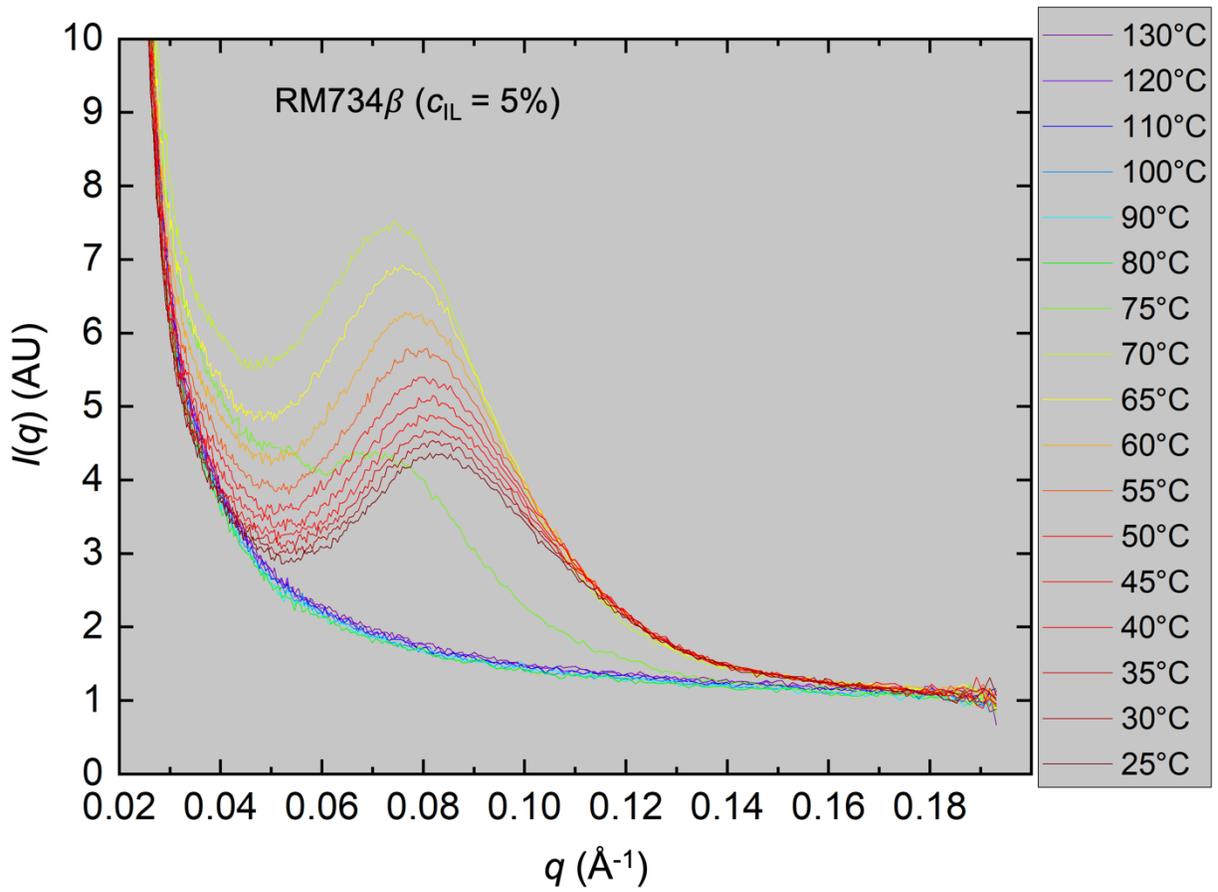

*Figure S7:* SAXS scans of RM734$_\beta$/$c_{IL}$ = 5 wt% mixture.

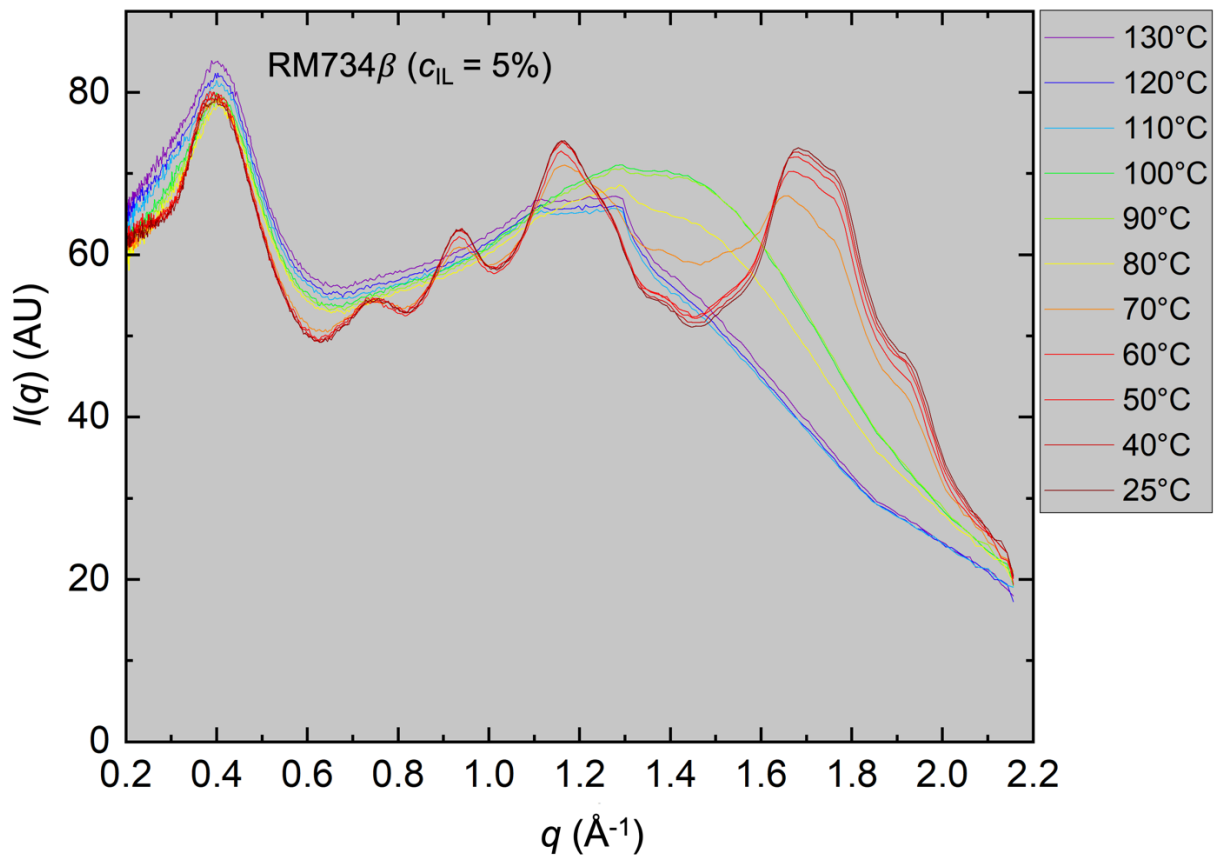

*Figure S8:* WAXS scans of RM734$_β$/$c_{IL}$ = 5 wt% mixture.

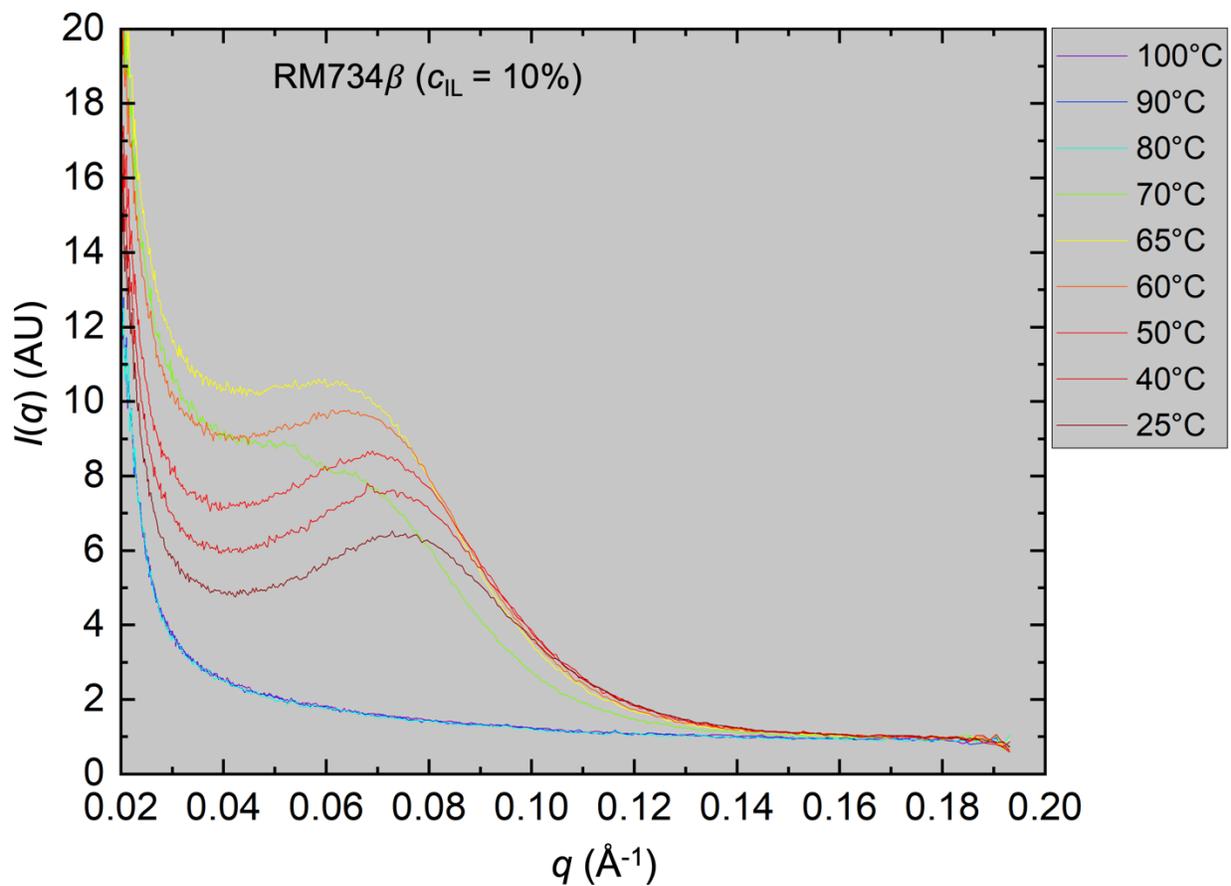

*Figure S9:* SAXS scans of RM734$_\beta$/$c_{IL}$ = 10 wt% mixture.

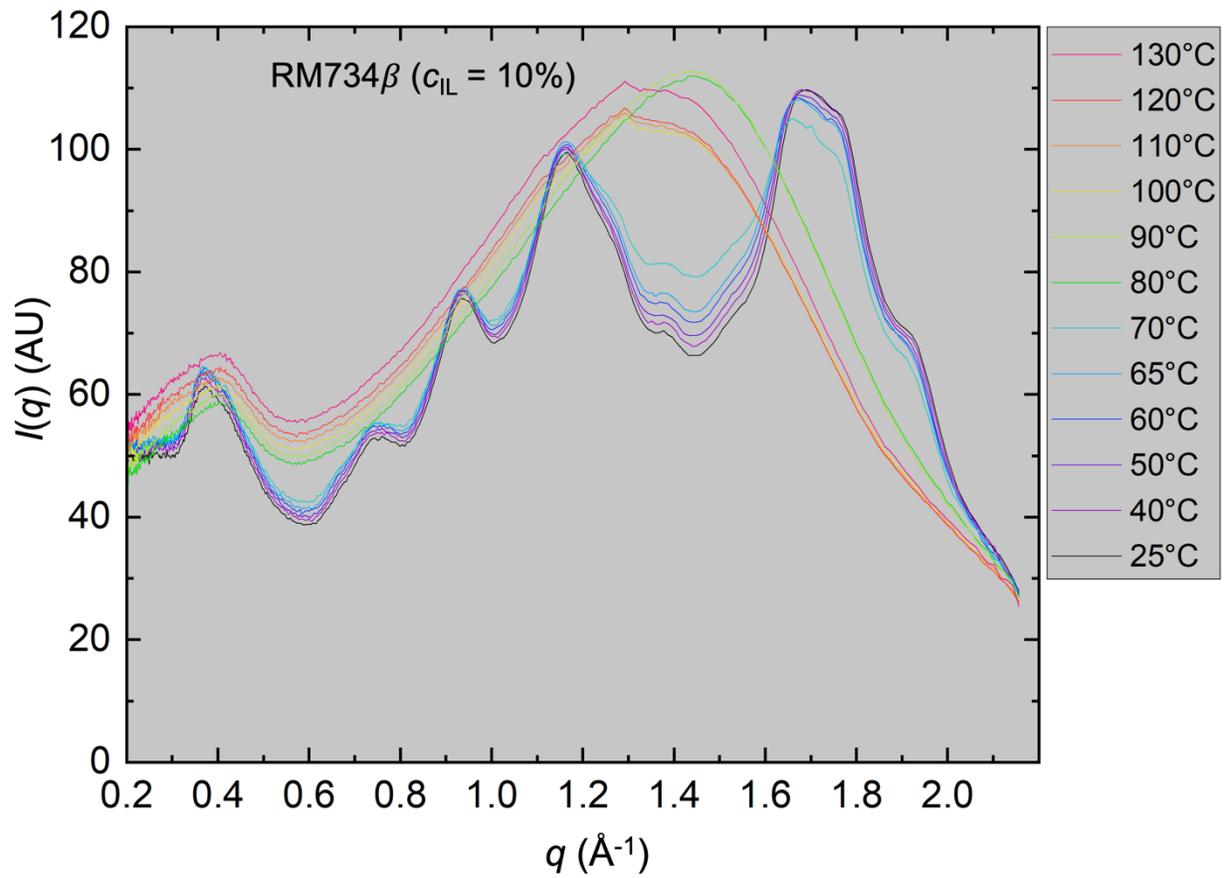

*Figure 10:* WAXS scans of RM734$_\beta$/$c_{IL}$ = 10 wt% mixture.